\documentclass[preprint,12pt,authoryear]{elsarticle} 
\usepackage{amsmath, amssymb, amscd, amsthm, amsfonts}
\usepackage{ulem}
\usepackage{caption}
\usepackage{subcaption}
\usepackage{blindtext}
\usepackage{graphicx}
\usepackage{hyperref}
\usepackage{s tackengine}
\usepackage[utf8]{inputenc}
\usepackage{natbib}
\usepackage{float}
\usepackage{array}
\usepackage{mathtools}
\usepackage{tabularx}
\usepackage{array}
\usepackage{multirow}
\usepackage{xltabular}
\usepackage{changepage}
\usepackage{xcolor}

\journal{Infectious Disease Modelling}

\begin{document}

\begin{frontmatter}



\title{Deterministic epidemic models overestimate the basic reproduction number of observed outbreaks}


\author[inst1]{Wajid Ali}

\affiliation[inst1]{organization={Department of Mathematical Sciences, University of Liverpool},
            addressline={Peach Street}, 
            city={Liverpool},
            postcode={L69 7ZX}, 
            state={England},
            country={United Kingdom}}

\author[inst1]{Christopher E. Overton}
\author[inst2]{Robert R. Wilkinson
}
\author[inst1]{Kieran J. Sharkey\corref{cor1}}
\ead{kjs@liverpool.ac.uk}

\affiliation[inst2]{organization={Department of Applied Mathematics, Liverpool John Moores University},
            addressline={ Byrom Street}, 
            city={Liverpool},
            postcode={L3 5UX}, 
            state={England},
            country={United Kingdom}}

\cortext[cor1]{Corresponding author}
\begin{abstract}
The basic reproduction number, $R_0$, is a well-known quantifier of epidemic spread. However, a class of existing methods for estimating $R_0$ from incidence data early in the epidemic can lead to an over-estimation of this quantity.
In particular, when fitting deterministic models to estimate the rate of spread, we do not account for the stochastic nature of epidemics and that, given the same system, some outbreaks may lead to epidemics and some may not.
Typically, an observed epidemic that we wish to control is a major outbreak.
This amounts to implicit selection for major outbreaks  which leads to the over-estimation problem. We formally characterised the split between major and minor outbreaks by using Otsu's method which provides us with a working definition.
We show that by conditioning a `deterministic' model on major outbreaks, we can more reliably estimate the basic reproduction number from an observed epidemic trajectory.
\end{abstract}




\end{frontmatter}

\section{Introduction}
A new, emerging infectious disease can potentially spread around the world within days or weeks, as observed during COVID-19~\citep{Carvalho2021},  and swine flu~\citep{Coker2009}. During the early phase of an epidemic, estimation of key epidemiological parameters helps us to estimate its future behaviour including the rate of spread, final size, and the requirements for effective control. In these early stages, epidemics typically exhibit exponential growth.

The basic reproduction number, $R_0$, is  the average number of secondary infections per primary infection in an otherwise susceptible population \citep{Dietz1993,Heesterbeek1996,Heffernan2005}. 
The basic reproduction number has been shown to have important implications relating to the final epidemic size \citep{Andreasen2011} and requirements for control \citep{Lipsitch2003}.  It is also directly related to the early growth rate ($r$) of epidemics \citep{Lipsitch2003,Ma2020} and both of these quantifiers are used for predicting the fate of outbreaks. 
That is, when $R_0$ is greater than $1$ (or $r$ is positive), then the introduction of an infected individual into a susceptible population may lead to a major epidemic. Given the random nature of infection processes, stochastic models are the natural choice to model epidemics~\citep{Whittle1955,Bailey1975, Britton2019}, and methods such as maximum likelihood can be used to estimate $R_0$, taking into account this inherent randomness~\citep{Becker1999,Britton2019,ma2014}.

 Although epidemics are stochastic processes, it is sometimes convenient to use a deterministic approach such as the Kermack-Mckendrick SIR model~\citep{Kermack1927, Kermack1932}, SIS model~\citep{Lajmanovich1976}, SEIR Model~\citep{Anderson1992} or the exponential or logistic growth curves~\citep{Chowell2006}   to understand and predict them. Unlike their stochastic counterparts \citep{Whittle1955, Bailey1975, Britton2019}, these models guarantee an epidemic when $R_0>1$ \citep{Dietz1993, Kermack1927}. Such models have been used to estimate epidemic parameters by fitting them to real epidemic data, for example,  influenza~\citep{Chowell2016}, cholera~\citep{Pourabbas2001}, and COVID-19~\citep{Metelmann2021}. We refer to \citep{ma2014} and \citep{Ma2020} for more details on estimating early growth rates and the basic reproduction number from real data. 

These classic models can be valuable but may lead to an overestimation of the basic reproduction number~\citep{Breban2007,Chowell2017,Green2006,Keeling2000}.
Generally, uncertainty in the estimation of parameters may arise due to noise in the data and/or the underlying assumptions for building models~\citep{Chowell2017,Ferrari2005}. 
However, here we observe that there is also a fundamental bias in the deterministic models which occurs because they do not capture the stochastic effects in the early phases of an outbreak and, in particular, do not distinguish the possibility of stochastic fade-out when $R_0$ is greater than 1 \citep{Whittle1955, Bailey1975}.
Similar issues with deterministic models and stochastic fade-out have been explored in \citep{Overton2022} in the context of steady-state solutions to the SIS model. 

By reducing them to a simple birth-death process, we show that SIR and SIS deterministic models implicitly average over both major and minor outbreaks during their early phases; that is both extinct and extant trajectories are included in the average behaviour.
However, an observed epidemic is necessarily a major outbreak and therefore corresponds to an implicit conditioning on major outbreaks.
This leads deterministic SIR and SIS models to overestimate  the basic reproduction number, $R_0$, when they are fitted to epidemic data which we illustrate in the next section. 
This is more pronounced when the probability of minor outbreaks is large; i.e. when we have a small number of initial infections or when $R_0$ is close $1$.  
 In Section~\ref{sec:conditioned_method} we consider a birth-death process conditioned on major outbreaks which we approximate by conditioning on non-extinction   \citep{Kot2001,Kendall1948a,Kendall1948b}. This better-describes a typical major outbreak and we show that it performs well in removing the bias from the estimation of $R_0$. 

\section{Estimation of \texorpdfstring{$R_0$}{R0} using standard deterministic models}\label{sec:model_framework}
\noindent Consider an infectious disease that is spread via contact between susceptible and infected individuals  in a well-mixed homogeneous population. Let $\tau$ be the rate at which a single individual infects a susceptible individual during its infectious period and let $i(t)$ and $s(t)$ denote the infectious and susceptible populations respectively. We consider an infectious disease with no latent period and  suppose that infection  occurs according to a Poisson process with rate $\tau si$. Similarly, we assume removal (or recovery) occurs according to a Poisson process with rate $\gamma i$ where $\gamma$ is the rate of removal/recovery of a single individual. This can be applied to infections that produce no long-term immunity (SIS or SIRS) or permanent immunity (SIR).

\begin{figure}
  \begin{subfigure}{0.32\textwidth}
       \caption{Averages of SIR process}
     \includegraphics[width=\textwidth]{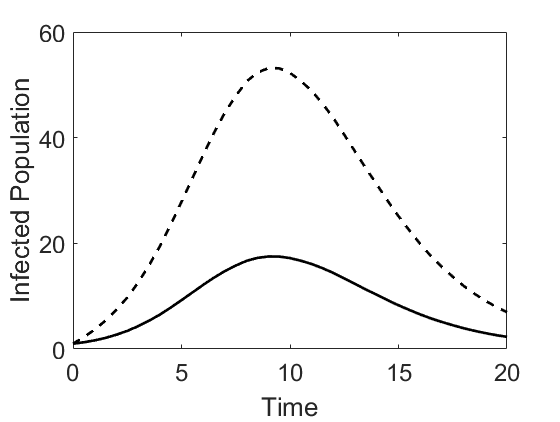}
     \label{fig:sir_sims}
 \end{subfigure}
 \hfill
 \begin{subfigure}{0.32\textwidth}
 \caption{Averages of SIS process}
     \includegraphics[width=\textwidth]{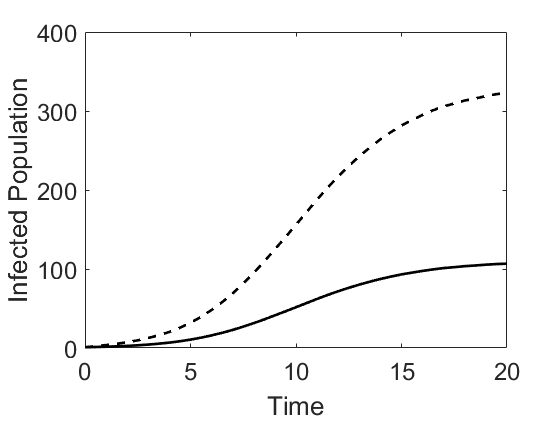}
     \label{fig:SIS_sims}
 \end{subfigure}
  \begin{subfigure}{0.32\textwidth}
   \caption{Averages of BD process}
     \includegraphics[width=\textwidth]{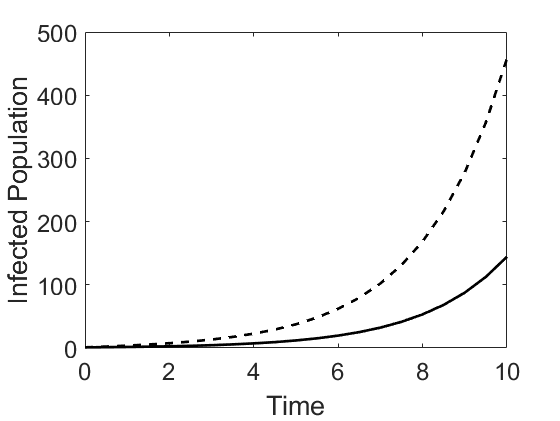}
     \label{fig:BD_sims}
 \end{subfigure}  
 \caption{The average of 10,000 simulations (solid line) and those conditioned on major outbreaks (dashed line) for (a) the SIR process, (b) the SIS process and (c) the simple Birth-Death (BD) process. In each case, $\beta=1.5$, $\gamma=1$, $N=1000$ and the initial number infected is $i_0=1$.}
 \label{fig:sims}
\end{figure}

In a sufficiently large population, the early phase of the epidemic behaves like a simple birth-death (BD) process. This can be seen from the infection rates $\tau si$; for a total population of size $N$ initiated with a single infected ($i_0=1$) in an otherwise susceptible population, the initial infection rates for $i=1,i=2,i=3,\dots$ infected individuals are $\tau(N-1), 2\tau(N-2), 3\tau(N-3),\dots$ respectively. These are approximately $i\tau N =\beta i$ because the susceptible population is approximately $N$ \citep{Renshaw1993}. So, the early phases of the infection dynamics are approximated by a simple birth-death process with individual birth rate $\beta$ and individual death rate $\gamma$. For comparison, both SIS and SIR processes and (their approximation) the simple BD process are summarised in Table~\ref{tab:sis_sir_bd}, and the time series curves of the infected populations are illustrated in Fig.~\ref{fig:sims}. 
\begin{table}
\centering
\begin{tabular}{ | c | c| c|c | } 
  \hline
    Event   & SIS & SIR & BD \\ 
  \hline
 Infection  &$(s,i) \xrightarrow{\tau si}  (s-1,i+1)$ & $(s,i) \xrightarrow{\tau si}  (s-1,i+1)$ & $i\xrightarrow{\beta i}i+1$\\
  \hline
Recovery  &$(s,i) \xrightarrow{\gamma i} (s+1,i-1)$ & $(s,i) \xrightarrow{\gamma i}  (s,i-1)$ & $i\xrightarrow{\gamma i} i-1$\\
  \hline
\end{tabular}
\caption{State transitions in the Susceptible-Infected-Susceptible (SIS), Susceptible-Infected-Recovered (SIR) and the simple Birth-Death (BD) processes. }
\label{tab:sis_sir_bd}
\end{table}

The expected number of  infected individuals (denoted by $\langle i\rangle$) in the simple BD process at a given time $t$ can be derived from the master equation for the process. Let the probability that there are $i$ infected individuals at time $t$ be denoted by $p_i(t)$, where $i\in\{0,1\dots\}$. Thus, the master equation for the simple birth-death process is given by:
\begin{equation}\label{equ:disease_birth-death}
    \frac{d}{dt}p_i(t)=\beta(i-1)p_{i-1}(t)-(\beta+\gamma)ip_i(t)+\gamma (i+1)p_{i+1}(t).
\end{equation}
From this, the rate of change of the expected infectious population is \citep{Feller1939, Kendall1948a}: 
\begin{equation}\label{equ:exp_BD}
  \frac{d}{dt}	\langle i 	\rangle=\sum_i i\frac{d}{dt}p_i(t)= (\beta-\gamma)	\langle i 	\rangle.
\end{equation}
 
 This has the same form as the deterministic simple BD model given in Table~\ref{tab:det_models}, so the deterministic BD model describes the average of all stochastic realisations of the stochastic BD process. This connection is well-known but unusual, although similar connections can be established between the stochastic and  deterministic SIS and SIR models under some closure approximations~\citep{Sharkey2015,Kiss2017} and in limiting cases~\cite{Kurtz1970}.
 \begin{table}
     \centering
     \begin{tabular}{|c|c|c|}
   \hline
 SIS & SIR & BD \\ 
\hline
{$\!\begin{aligned}  
               \frac{dS}{dt}&=-\tau SI+\gamma I \\  
               \frac{dS}{dt}&=\tau SI-\gamma I
\end{aligned}$}
&
{$\!\begin{aligned}  
               \frac{dS}{dt}&=-\tau SI \\  
               \frac{dS}{dt}&=\tau SI-\gamma I\\
               \frac{dR}{dt}&=\gamma I
               
\end{aligned}$} 
&
{$\!\begin{aligned}  
\frac{dI}{dt}=\beta I-\gamma I
\end{aligned}$}\\
\hline                               
     \end{tabular}
     \caption{Equations for deterministic SIS, SIR and BD models where $S$, $I$, $R$ are the sizes of the classes of susceptible, infected and removed respectively. The parameters $\tau$ and $\gamma$ are the transmission and recovery (or removal) rates. Here, for convenience we have used the same parameter notation as for the stochastic models. }
     \label{tab:det_models}
     \end{table}

 Moreover, the deterministic BD model (and equivalently Equation~\ref{equ:exp_BD}) describes the expected early deterministic dynamics of both SIS and SIR processes when $N$ is large. This is because (see Table~\ref{tab:det_models}) the deterministic SIS and SIR (and SIRS) models have the following equation for the infectious  population \citep{Kermack1927}: 
 \begin{equation}\label{equ:det_dynamics}
\frac{dI}{dt}=\tau SI- \gamma I,
 \end{equation}
 where we use capital letters $I$ and $S$ for denoting the number of infected and susceptible individuals in deterministic models. This reduces to the form of Equation \ref{equ:exp_BD} under the same approximation as we applied to the stochastic models (i.e. $S\approx N$ with $\beta=\tau N$) and so the expected behaviour of the stochastic SIR and SIS models is approximated by the deterministic SIR and SIS models, and by the BD model  in the early stages.

  The equivalence of the deterministic models to the expected value of the stochastic  models and their derivation from the master equation  tells us that the deterministic epidemic models approximate an averaging over all epidemic outcomes~\citep{Overton2022,Kurtz1970}. Crucially this averaging is over both major and minor outbreaks. However, a real epidemic of interest is a major outbreak and this therefore represents conditioning on major outbreaks.
 
\begin{figure}
 \begin{subfigure}{0.49\textwidth}
  \caption{Final sizes in SIR process}
     \includegraphics[width=\textwidth]{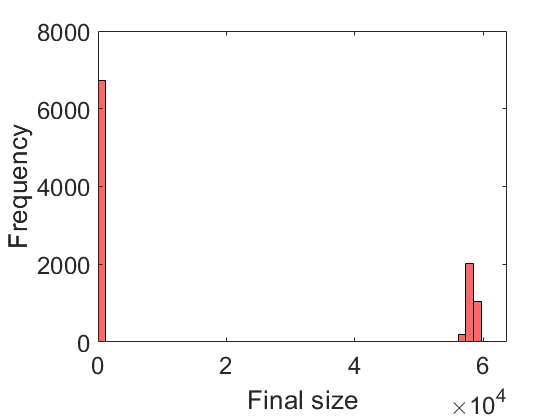}
     \label{fig:final_sizes}
 \end{subfigure}
 \hfill
  \begin{subfigure}{0.49\textwidth}
  \caption{Infection events in SIS process}
 \includegraphics[width=\textwidth]{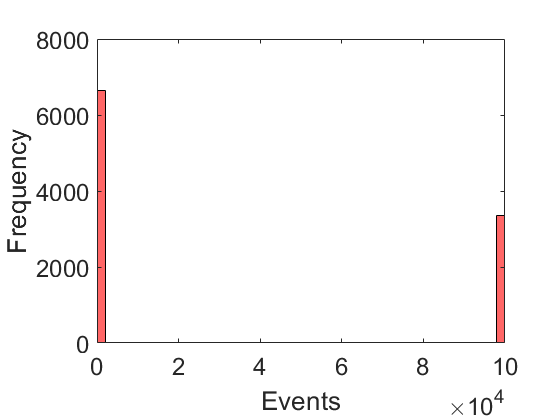}
\label{fig:SIS_events}
 \end{subfigure}
 \caption{(a) The final size distribution obtained from $10,000$ stochastic simulations of SIR dynamics and (b) the distribution in the number of (infection and recovery) events for 10,000 SIS simulations capped at $2N$ events. In both sets of simulations, $N=10,000$, $\beta=1.5$, $\gamma=1$, and the initial number infected is $i_0=1$.}
 \end{figure}

 Fig. \ref{fig:final_sizes} shows the distribution of final sizes \citep{Andreasen2011} of an SIR process when initiated with a single infected individual. The bimodal nature of this distribution tells us that that a group of realisations generate major outbreaks while others go extinct in the early phase. Fig.~\ref{fig:SIS_events} shows similar bimodal behaviour for SIS dynamics where here the process is run until either extinction or until $2N$  events have occurred. Although the split between major and minor outbreaks is usually obvious, it is not well-defined in finite populations. Throughout this paper we choose to formally characterise the split between major and minor outbreaks by using Otsu's method \citep{Otsu1979} which gives a threshold value for clustering bimodal histograms and provides us with a working definition. 
 
To illustrate the issue of over-estimation, we performed least-squares fits  of both deterministic SIR and SIS models to simulated major outbreaks of both types to estimate the transmission rate $\beta$, assuming that we know $\gamma$ (here $\gamma=1$). $R_0$ is then calculated from $R_0=\beta/\gamma$ \citep{Keeling2000,Driessche2017}. 

For all least-squares fits in the paper, epidemic incidence data was constructed by taking the number of infection events between one time step and the next: $j_t=c(t+\delta t)-c(t)$ where $c(t)$ is the cumulative number of infection events up to time $t$ given by $c(t)=N-S(t)$ in the SIR case. We minimise
\begin{equation}
SSE=\sum_{n=0}^{T/\delta t}(J_{n\delta t}-j_{n\delta t})^2
\nonumber
\end{equation}
where $J_t=C(t+\delta t)-C(t)$ and $C(t)$ is the cumulative number of infections in the corresponding deterministic model \citep{ma2014}. Here, $T$ is the upper end of the fit window and we use $\delta t=0.1$. Minimisation was performed using the Nelder-Mead simplex algorithm as implemented by the fminsearch function in MATLAB ver. R2023a.

\begin{figure}
  \begin{subfigure}{0.49\textwidth}
       \caption{SIR model fits}
     \includegraphics[width=\textwidth]{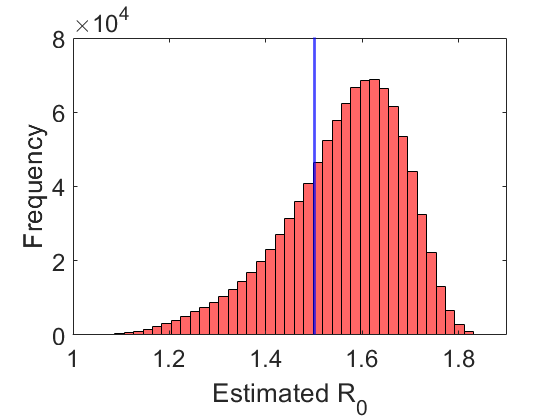}
     \label{fig:SIR_R0}
 \end{subfigure}
 \hfill
 \begin{subfigure}{0.49\textwidth}
      \caption{SIS model fits}
     \includegraphics[width=\textwidth]{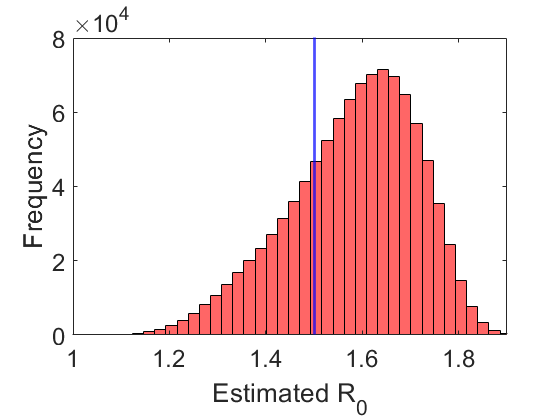}
     \label{fig:SIS_R0}
 \end{subfigure} 
 \caption{The distributions of $R_0$  estimated by performing least squares fits of (a) the SIR model and (b) the SIS model to $1$ million major outbreaks of each type generated by taking  $\beta=1.5$, $\gamma=1$,  $N=100,000$ and the initial number infected $i_0=1$. The blue lines in both plots represent the actual value of $R_0$ used to generate the simulated outbreaks. The fitting window for both histograms is $t\in [0, 10]$.}
 \label{fig:mot_plots}
\end{figure}

The resulting distributions in $R_0$ values are shown for both SIR (Fig.~\ref{fig:SIR_R0}) and SIS (Fig.~\ref{fig:SIS_R0}). Both distributions clearly show an upward bias in the estimates with respect to the ``true''  value of $R_0=1.5$ used to generate the stochastic simulations. 
Thus, fitting mean-field SIS and SIR models (or the simple BD model) to major real outbreak data will have a tendency to overestimate the transmission rate and overestimate $R_0$. 

 To resolve this and obtain a deterministic model that we can fit to the early phases of epidemics to determine $R_0$, we need to account for the implicit conditioning on major outbreaks. Since the early stages of SIS and SIR dynamics are well-approximated by the simple BD process, we try to obtain a simple BD process which is conditioned on major outbreaks.
 \section{Estimation of \texorpdfstring{$R_0$}{R0} using a conditioned BD 
 model}\label{sec:conditioned_method}
We wish to calculate the conditional probability $P(\textrm{$i$ infected at time $t|$ major outbreak})$ for the simple BD process. Due to the ambiguity in defining a major outbreak and to make analytic progress, we approximate this probability by:
\begin{equation*}
   P(\textrm{$i$ infected at time $t |$ major outbreak}) \approx q_i(t)
\end{equation*}
   where
\begin{equation*}
   q_i(t)= P(\textrm{$i$ infected at time $t | i\neq 0$ at time $t$})
\end{equation*}
for $i\in\{1\dots\}$.
   This is given by  \citep{Kot2001}
\begin{equation*}
   q_i(t)=\frac{p_i(t)}{1-p_0(t)}
\end{equation*}
 where, with reference to Equation \ref{equ:disease_birth-death}, the probability that the disease has died out at time $t$, is denoted by $p_0(t)$. Differentiating gives
$$ \frac{d}{dt}q_i(t)=\frac{\frac{d}{dt}p_i(t)}{1-p_0(t)}+\frac{p_i(t)}{\left[1-p_0(t)\right]^2}\frac{d}{dt}p_0(t),\;\;\;\textrm{for}\;\;\; i=1,2,\dots.$$
Using the expression for $dp_i(t)/dt$ given in Equation \ref{equ:disease_birth-death} when $i\neq 0$, and using $dp_0(t)/dt=\gamma p_1(t)$ and $q_i(t)=p_i(t)/(1-p_0(t))$, we reach the following model~\citep{Kot2001}:
\begin{equation}\label{equ:conditioned_birth-death}
    \frac{d}{dt}q_i(t)=\beta(i-1)q_{i-1}(t)-(\beta+\gamma)iq_i(t)+\gamma (i+1)q_{i+1}+\gamma q_1q_i(t).
\end{equation}

We now derive an equation for the rate of change of the expected number of infected individuals at time $t$: 
\begin{equation*}
\langle i 	\rangle_c(t)=\sum_{i=1}iq_i(t),	
\end{equation*}
where the subscript \(c\) indicates that this is the expected value in the conditioned process. Differentiating with respect to time and substituting for $dq_i(t)/dt$ from Equation \ref{equ:conditioned_birth-death} gives:
\begin{equation*}
    \frac{d}{dt}	\langle i 	\rangle_c=\beta\sum_{i=1}i(i-1)q_{i-1}(t)-(\beta+\gamma)\sum_{i=1}i^2q_i(t)+\gamma\sum_{i=1} i(i+1)q_{i+1}+\gamma q_1	\langle i 	\rangle_c.
\end{equation*}
The summation in the first term on the right-hand-side can be written as:
\begin{equation*}
\sum_{i=1}i(i-1)q_{i-1}(t)=\sum_{k=0}(k+1)(k)q_{k}(t)=\sum_{i=1}i^2q_{i}(t)+\sum_{i=1}iq_{i}(t), 
\end{equation*}
and the summation in the third term on the right-hand-side can be written as: 
\begin{equation*}
\sum_{i=1}i(i+1)q_{i+1}(t)=\sum_{k=2}(k-1)(k)q_{k}(t)=\sum_{i=2}i^2q_{i}(t)-\sum_{i=2}iq_{i}(t).
\end{equation*}
This leads to:
\begin{equation}\label{equ:expected_CBD}
 \frac{d}{dt}	\langle i 	\rangle_c=(\beta-\gamma)	\langle i 	\rangle_c+ \gamma	\langle i 	\rangle_c q_1
\end{equation}
with the cumulative number of infections given by
\begin{equation*}
\frac{dC}{dt}=\beta	\langle i 	\rangle_c+ \gamma	\langle i 	\rangle_c q_1. 
\end{equation*}
 This system is not closed because we have one extra variable, $q_1$, which is the probability of single infection in the conditioned process. However, using the exact solution of Equation \ref{equ:disease_birth-death}  \citep[Chapter~3]{Renshaw1993,Kot2001}, the exact expression for  $q_1$ is given by
\begin{equation*}\label{equ:q1}
    q_1(t)=\frac{p_1(t)}{1-p_0(t)}
\end{equation*}
where
$$p_0(t)=\left\{ 
    \begin{array}{lcl}
       \left[\frac{\beta t}{1+ \beta t}\right]^{i_0}&\textrm{if}  & \beta=\gamma \\
       \left[\frac{\gamma(e^{rt}-1)}{\beta e^{et}-\gamma}\right]^{i_0} &\textrm{if}  &  \beta \neq \gamma
    \end{array}
    \right. $$
and 
$$p_1(t)=\left\{ 
    \begin{array}{lcl}
i_0\left[\frac{\beta t}{1+ \beta t}\right]^{i_0-1}\frac{1}{(1+\beta t)^2} &\textrm{if}& \beta=\gamma \\
i_0 \alpha^{i_0-1}\left[(1-\alpha)(1-\phi)\right] &\textrm{if}  &  \beta \neq \gamma.
    \end{array}
    \right.$$
Here $i_0$ is the initial number of infected individuals and 
$$\alpha=\frac{\gamma(e^{rt}-1)}{\beta e^{rt}-\gamma},\;\;\;\; \phi=\frac{\beta(e^{rt}-1)}{\beta e^{rt}-\gamma},$$ 
where $r=\beta-\gamma$. 

\begin{figure}
   \begin{subfigure}{0.49\textwidth}
    \caption{Models vs simulations}
     \includegraphics[width=\textwidth]{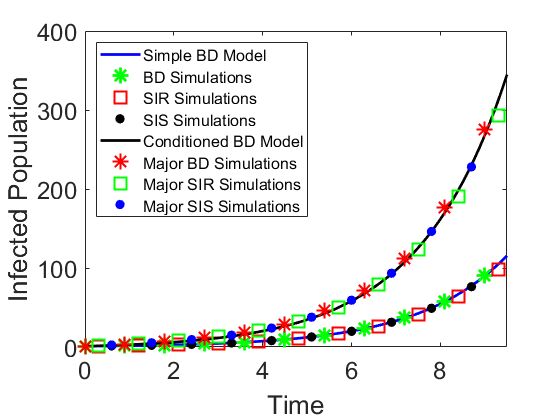}
     \label{fig:comp_all}
 \end{subfigure}
 \begin{subfigure}{0.49\textwidth}
      \caption{Conditioned BD model fits}
     \includegraphics[width=\textwidth]{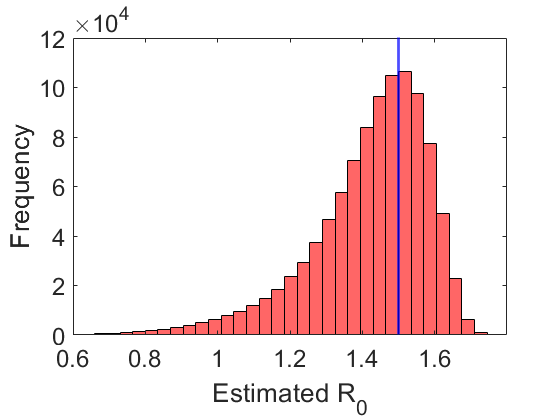}
     \label{fig:CBD_R0}
 \end{subfigure}
 \caption{(a) The deterministic conditioned and unconditioned BD models together with averages of 100,000 conditioned and unconditioned BD, SIS and SIR simulations. (b) Estimates of $R_0$ by least-square fits of the conditioned BD Model~(Equation \ref{equ:expected_CBD}) to 1 million major outbreaks. The fitting window used is $t\in[0, 9.5]$ which is  determined so that the simple BD model is within $1\%$ of  the SIR infectious time series. For both subplots, $i_0=1$, $\beta=1.5$, $\gamma=1$ and  $N=100,000$.}
\end{figure}

\begin{figure}
\begin{subfigure}{0.49\textwidth}
    \caption{Conditioned BD Model}
    \includegraphics[width=\textwidth]{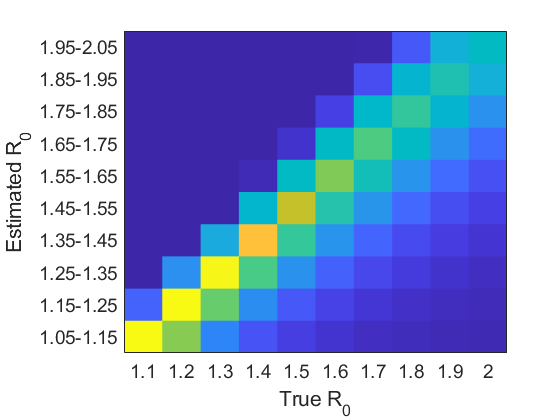}

    \label{fig:CI_CBDSIR}
\end{subfigure}
\hfill
\begin{subfigure}{0.49\textwidth}
 \caption{SIR Model}
    \includegraphics[width=\textwidth]{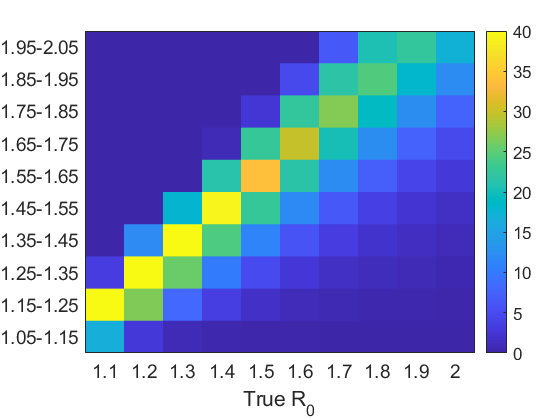}
   
    \label{fig:CI_SIR}
\end{subfigure}
\label{fig:heatmaps}
 \caption{The distribution of estimated $R_0$ values by fitting (a) the conditioned BD model and (b) the SIR model. Both these models are fitted to 1 million simulated major SIR outbreaks per $R_0$ value.  The horizontal axis corresponds to the ``true'' values of $R_0$ used to generate the simulations. The vertical axis corresponds to the histograms of estimated $R_0$ (similar to Figure~\ref{fig:CBD_R0}). The intensity of the color indicates the percentage of estimates in each interval.}
 \label{fig6}
\end{figure}

As discussed in Section \ref{sec:model_framework}, the early phases of the SIS and SIR epidemic models are well-approximated by the simple BD process. Similarly the conditional (SIS and SIR) epidemic processes are also well-approximated by the conditional stochastic BD process. Conditioned and non-conditioned processes lead to different average trajectories (Fig.~\ref{fig:comp_all}) and it can be seen  that the new conditioned BD model (Equation \ref{equ:expected_CBD}) accurately describes the early part of the SIR, SIS and BD expected behaviour when  these processes are conditioned on major outbreaks. This model resolves the issue of upward bias which we saw in Fig.~\ref{fig:SIR_R0}. When it is fitted to  major outbreaks of SIR epidemics, we can see in Fig.~\ref{fig:CBD_R0} that the distribution of estimated $R_0$ values is centred around the true value. Fig. \ref{fig:CI_CBDSIR} explores the parameter space for small $R_0$ more fully and shows a significantly improved estimate of $R_0$ when compared with the standard SIR model (Fig.~\ref{fig:CI_SIR}). The fitting window for Fig.~\ref{fig:CBD_R0} and Fig.~\ref{fig6} is the time interval $t\in [0\,\, T]$ where we determined $T$ to be the time point at which the simple BD approximation gives a 1\% error with respect to the infectious time series in the SIR process.

\section{Discussion}\label{sec:results_and_discussion}
\noindent One method for obtaining the basic reproduction number ($R_0$) from real data is to fit deterministic models to epidemic data~\citep{Breban2007,Chowell2017,Driessche2017,Green2006, Keeling2000,Ma2020}. Here we showed that some of these deterministic models are fundamentally biased in their estimation of $R_0$~\citep{Breban2007,Chowell2017}. We showed this explicitly in the context of SIS and SIR epidemic dynamics (Fig. \ref{fig:SIR_R0} and \ref{fig:SIS_R0}) where we can make some analytic progress. This also applies directly to SIRS models as well \citep{Mena-Lorcat1992}.

The early phases of both the stochastic SIS and SIR epidemic processes and the deterministic SIS and SIR models reduce to a simple birth-death process (see, for example, \citep{Renshaw1993}), and so the early behaviour of the deterministic models describes the expected behaviour of all possible epidemic realisations. The over-estimation of $R_0$ occurs because a real epidemic of interest is necessarily a major outbreak, representing an implicit conditioning on major outbreaks. This overestimation arises when deterministic epidemic models are initialised with very few initial infected individuals ($i_0$) and/or where $R_0$ is very close to 1 which leads to the probability of a minor outbreak (given by $(1/R_0)^{i_0}$\citep{Whittle1955}) to be significant. Once the epidemic is underway and the deterministic models can be initialised with sufficient infected individuals to make the probability of minor outbreaks negligible, then the overestimation problem does not arise. 

We resolved the issue of overestimation by developing a simple birth-death model with conditioning on major outbreaks. To make analytic progress we approximated conditioning on major outbreaks by  conditioning against extinction (Equation \ref{equ:expected_CBD})  \citep{Kot2001,Kendall1948a,Kendall1948b}. We made use of the analytic solution of the simple birth-death master equation and showed that fitting this model to the early stages of epidemic outbreaks resolves the issue of overestimation of $R_0$ (Fig. \ref{fig:CBD_R0} and \ref{fig:CI_CBDSIR}). Similar models were used for approximating the average phylogenic lineages~\citep{Harvey1994} and for correcting a similar bias in deterministic coalescent models~\citep{Stadler2015}. 

It is expected that similar issues would apply to other epidemic models such as SEIR type models~\citep{Anderson1992}  and models that do not rely on Poisson processes~(\citep{Kenah2011,KhudaBukhsh2019}). It would be of interest to determine the extent of this error in these scenarios. For this, we currently lack the ingredients that we made use of which are the approximation of SIS and SIR dynamics by the simple birth-death (BD) process, the analytic solution of the BD process and the equivalence of the deterministic BD and the average stochastic BD processes. Nevertheless, this implicit conditioning on major outbreaks seems likely to cause problems with these models as well. The approximation to the conditioned SIS model proposed in \citep{Overton2022} can also be extended to cover transient dynamics and to model more complex epidemiological structures. While the validity of using a deterministic description early in an epidemic is questionable due to the stochastic fluctuations of the infected population, the main fluctuation is the one between major and minor outbreaks. Our work emphasises the importance of not defining the initial conditions of standard deterministic models too early in the epidemic unless the models have conditioning against minor outbreaks.
 
\section*{Acknowledgements}
This project has received funding from the European Union’s Horizon 2020 research and innovation programme under grant agreement No 955708.

\bibliographystyle{elsarticle-harv} 
\bibliography{library}
\end{document}